\input amstex
\documentstyle{amsppt}
\TagsOnRight   \magnification \magstep1 \baselineskip 18 pt \pageheight{7.5in}
\pagewidth{5.42in}

\heading Unified Field Theory From Enlarged Transformation Group.  The Covariant
Derivative for Conservative Coordinate
Transformations and Local Frame Transformations\\
Edward L. Green \\
North Georgia College \& State University  \endheading

\vskip .5in

\par \phantom{D} \par
\subhead Abstract
\endsubhead
\flushpar Pandres has developed a theory in which the geometrical structure of a real
four-dimensional space-time is expressed by a real orthonormal tetrad, and the group of
diffeomorphisms is replaced by a larger group called the conservation group. This paper
extends the geometrical foundation for Pandres' theory by developing an appropriate
covariant derivative which is covariant under all local Lorentz (frame)
transformations, including complex Lorentz transformations, as well as conservative
transformations. After defining this extended covariant derivative, an appropriate
Lagrangian and its resulting field equations are derived.  As in Pandres' theory, these
field equations result in a stress-energy tensor that has terms which may automatically
represent the electroweak field.  Finally, the theory is extended to include 2-spinors
and 4-spinors.  Note: This article was published by the International Journal of
Theoretical Physics (2009) {\bf 48}: 323-336.  DOI: 10.1007/s10773-008-9805-z.  The
original publication is available at http://www.springer.com/physics/journal/10773 .

\vskip 0.7in

\flushpar Keywords:  Field theory, transformation groups, covariant derivatives,
lagrangians, field equations, spinors

\par \phantom{D}
\flushpar PACS:  12.10-g, 4.20.Fy, 4.20.Gz

\newpage

{\bf  1. Introduction.}
\par \phantom{D} \par

Previously a theory has been presented which exhibits many of the features required for
a unified field theory (Pandres 1981, 1984, Green and Pandres, 2003).  The main feature
is invariance under a group of transformations that is larger than the diffeomorphism
group.   We will consider a 4-dimensional space $X^4$ which have local coordinates
$x^\mu\;$ ($\mu=0,1,2,3$) and regard the tetrad $h^i_{\; \mu}\;$  (with $i=0,1,2,3$) as
the contracted product of the field variables $h^I_{\;\mu}$ and $L^i_I$ (defined
below). Under the enlarged group of transformations which is defined below, quantities
such as the tetrad may be path-dependent.  The values of $x^i$ are considered to be
inertial coordinates with metric $\eta_{ij} \equiv \text{diag}\bigl\{-1,1,1,1\bigr\}$
(we use the Einstein summation convention throughout this paper), and the metric tensor
is defined by $g_{\mu\nu}=\eta_{ij} h^i_{\;\mu} h^j_{\;\nu}$. When $h^i_{\;\mu}$ is a
function of $x^\mu$, i.e. path-independent, we may interpret $X^4$ as a 4-dimensional
(pseudo-)Riemannian manifold ${\Cal M}^4$ with metric $g_{\mu \nu}$. This is called the
{\it manifold interpretation}.

A Riemannian manifold is invariant under diffeomorphisms which for $x^\mu \to
x^{\bar{\mu}}$ satisfy the property $x^{\bar{\alpha}}_{\;\; ,\nu ,\mu} -
x^{\bar{\alpha}}_{\;\; ,\mu , \nu}=0$.  In the tetrad formulation, it is also invariant
under Lorentz transformations $L_{\bar{i}}^i$ which satisfy the condition
$\eta_{\bar{i}\bar{j}}=\eta_{i j}L_{\bar{i}}^i L_{\bar{j}}^j=
\text{diag}\bigl\{-1,1,1,1\bigr\}$.  The inverse of $L_{\bar{i}}^i$ will be denoted by
$L^{\bar{i}}_i$ and hence $L_{\bar{i}}^j L^{\bar{i}}_k = \delta^j_k$ and $L_i^{\bar{k}}
L_{\bar{j}}^i=\delta^{\bar{k}}_{\bar{j}}$.  Also $h_i^{\;\mu}$ is defined by the
requirement that at every point, $h_i^{\;\mu}h^i_{\;\nu}=\delta^\mu_\nu$. Under
diffeomorphisms on $x^\mu$ and Lorentz transformations on $x^i$, the Riemannian
manifold generated by $h^{\bar{i}}_{\;\bar{\mu}} = h^i_{\;\mu} L^{\bar{i}}_i
x^\mu_{\;\; ,\bar{\mu}} \;$ is the same as that generated by $h^i_{\;\;\mu}$.

When the frame transformation, $L^{\bar{i}}_j$ from one Latin system to another is
allowed to be a function of position (local), it is well-known that the transformation
from $x^i\rightarrow x^{\bar{i}}$ is not a diffeomorphism, i.e. the integrability
condition $L^{\bar{i}}_{j,k}-L^{\bar{i}}_{k,j}=0$ is not satisfied.   The value of
$x^{\bar{i}}$ will depend on the path in $x^i$ space and hence we cannot interpret the
$x^{\bar{i}}$ space as a manifold.  The special relativistic equation of a free
particle is $\dfrac{d^2 x^i}{ds^2}=0$.  Under local, non-diffeomorphic, Lorentz
transformations $L^{\bar{i}}_i$, this implies that $\dfrac{d^2
x^{\bar{i}}}{ds^2}=-L^{\bar{i}}_i L^i_{\bar{i},\bar{j}} \dfrac{dx^{\bar{i}}}{ds}
\dfrac{dx^{\bar{j}}}{ds}$ and thus we see that the $x^{\bar{i}}$ system is
non-inertial.

Therefore we have three spaces and convert between them using the field variables
$h^I_{\;\mu}$ and $L^i_I$, with $h^i_{\;\mu}=L^i_I h^I_{\;\mu}$.  Let $V_i$ be a vector
in the inertial space,
$$ \CD
  V_i  @> \quad \Lambda^i_I \quad >>  V_I  @> \quad h^I_{\;\mu} \quad
>>  V_\mu
\endCD
$$
We call the $x^i$ space the inertial space, the $x^I$ space the {\it internal space}
and the $x^\mu$ the {\it world space}. Analogous to the tetrad, we view $L^i_I$ as 4
internal vectors with inverse $L^I_i$ which satisfies $L^I_i L^i_J=\delta^I_J$ and
$L^I_j L^i_I= \delta^i_j$.  The fundamental fields are $L^i_I$ and $h^I_{\;\mu}$ since
$h^i_{\;\mu}$ is expressed by $h^i_{\;\mu}= L^i_I h^I_{\;\mu}$.  We will use capital
Latin indices such as $V^I$, $h^J_{\;\mu}$, etc. to denote the quantity in the internal
system.  Note that generally, $L^i_{I,J}-L^i_{J,I} \neq 0$.  We require that
$\eta_{_{IJ}} = \eta_{ij}L^i_I L^j_J = diag(-1,1,1,1)$. On the $x^I$ (internal) space,
we allow local (nonconstant) Lorentz transformations $L^{\hat{I}}_J$ while on the $x^i$
(inertial) space we allow only global (constant) Lorentz transformations, i.e.
$L^{\hat{i}}_{j,\mu}\equiv 0$. We will use the convention that when $L$ has a capital
subscript and a lowercase superscript or vice versa, that the $L$ represents the field
variable in the given system.  When both superscript and subscript are lowercase
letters, $L$ will represent a global Lorentz (frame) transformation and when both
superscript and subscript are capital letters, $L$ will generally represent a local
Lorentz (frame) transformation. When coordinates in the internal space $x^I$ are
changed $x^I\to x^{\hat{I}}$, then, in the new system, $h^{\hat{I}}_{\;\mu}=
h^I_{\;\mu} L^{\hat{I}}_I$ and $L^i_{\hat{I}}=L^i_I L^I_{\hat{I}}$.  Effectively, the
inertial space serves as a pregeometry upon which the richer geometry of the internal
space is founded and thence to the external (world) space geometry.

Since $h^I_{\;\mu}=h^i_{\;\mu}L^I_i$, then $g_{\mu\nu}=
\eta_{IJ}h^I_{\;\mu}h^J_{\;\nu}=\eta_{IJ}h^i_{\;\mu}L^I_i h^j_{\;\nu}L^J_j
=\eta_{ij}h^i_{\;\mu}h^j_{\;\nu}$.     Because the metric is unchanged, the field
variables $L^i_I$ do not affect the geometry of the manifold that is determined by
$h^i_{\;\mu}$.  If $h^i_{\; \mu ,\nu} - h^i_{\;\nu , \mu}=0$, then, in the manifold
interpretation, $X^4$ is a manifold with a vanishing curvature tensor, but this does
not imply that the internal space is flat, since $L^i_{I,J} - L^i_{J,I}$ may be
nonzero.  This may provide a framework for understanding the geometry of the vacuum.

For transformations on $X^4$, we consider a larger group of transformations which is
called the conservation group (Pandres, 1981). We say a transformation is {\it
conservative} if it satisfies the weaker condition
$$ x^\nu_{\;\; ,\bar{\alpha}}\bigl( x^{\bar{\alpha}}_{\;\; ,\mu ,\nu} - x^{\bar{\alpha}}_{\;\; ,\nu , \mu}
\bigr)=0 \quad . \tag{1}$$ The group of conservative transformations contains the
diffeomorphisms as a proper subgroup. In the Riemannian manifold interpretation we
regard $x^{\bar{\mu}}$ as anholonomic when $x^{\bar{\alpha}}_{\;\; ,\mu}$ is
non-diffeomorphic. We will use a semicolon to denote covariant differentiation with
Christoffel symbol $\Gamma^\alpha_{\mu\nu} = \frac12 \,g^{\,\alpha\sigma}\bigl(
g_{\sigma\mu ,\nu} + g_{\sigma\nu ,\mu}-g_{\mu\nu ,\sigma}\bigr)$. Let
$\tilde{V}^\alpha$ be a vector density of weight $+1$.  The conservation group of
transformations arises out of the requirement that a conservation law of the form
$\tilde{V}^\alpha_{\; ;\alpha}=0$ is preserved, i.e. $x^\alpha \to x^{\bar{\alpha}}$
being conservative implies that $\tilde{V}^{\bar{\alpha}}_{\; ;\bar{\alpha}}= 0$ as
well.   Dirac (Dirac, 1930) has remarked that "further progress lies in the direction
of making our equations invariant under wider and still wider transformations."   We
suggest that this enlargement of the transformation group results in a theory which
unifies gravity with the other forces.

As noted above, the field variables $L^I_i$ generally do not satisfy the integrability
condition: $L^I_{i,j}-L^I_{j,i} = 0$. We define {\it conservative Lorentz
transformations} by the requirement that
$$L^{I}_{\bar{I}}\bigl(L^{\bar{I}}_{J,I}-L^{\bar{I}}_{I,J} \bigr)=0 \quad . \tag{2}$$ Since
$L^{\bar{I}}_I$ is a Lorentz transformation the determinant of $L^{\bar{I}}_I$ is $\pm
1$ and hence the derivative of the determinant is zero. This implies that
$L^{I}_{\bar{I}} L^{\bar{I}}_{I,J} = 0$ and thus conservative Lorentz transformations
satisfy the condition $L^{I}_{\bar{I}} \, L^{\bar{I}}_{J,I}=0$.  Thus, with use of the
chain rule, we have $L^{\bar{I}}_J \text{ conservative } \iff L^{\bar{I}}_{J,\,
\bar{I}}=0 $. However, when we extend the group to complex Lorentz transformations (2)
must be used since the determinant of $L^I_{\bar{I}}$ is of the form $e^{i\theta(x)}$
and hence is not constant.   Although the only diffeomorphic Lorentz transformations
are global, there exist local (position-dependent) conservative Lorentz
transformations.  (The results of this paper do not depend on the concept of
conservative Lorentz transformations on $x^I$ space, but are included here for future
reference.)

We also recall that the Ricci rotation coefficient given by
$\gamma^\alpha_{\;\;\mu\nu}=h_i^{\;\alpha}h^i_{\;\mu ;\nu}$ is used to define the spin
connection.  However $\gamma^\alpha_{\;\;\mu\nu}$ is not a scalar under local Lorentz
transformations $L^i_{\bar{i}}$ since
$$\aligned h_i^{\;\alpha}h^i_{\;\mu ;\nu} &= h_i^{\;\alpha} \bigl( L^i_{\bar{i}}
h^{\bar{i}}_{\; \mu} \bigr)_{\, ; \nu} \\ &=h_i^{\; \alpha} L^i_{\bar{i}}
h^{\bar{i}}_{\;\mu ;\nu}+ h_i^{\; \alpha} h^{\bar{i}}_{\; \mu} L^i_{\bar{i}, \,\nu} \\
&=h_{\bar{i}}^{\;\alpha}h^{\bar{i}}_{\;\mu ;\nu} + h_i^{\;\alpha}h^{\bar{i}}_{\;\mu}
L^i_{\bar{i} \, ,\nu} \qquad .
\endaligned$$
In the manifold interpretation we see that the usual definition of
$\gamma^\alpha_{\;\mu\nu}$ results in a quantity that is not invariant under local
frame transformations.

{\bf Definition:}  When $h^i_{\; \mu} = L^i_I h^I_{\;\mu}$ is the tetrad used to define
a Riemannian manifold $\Cal M$, we define the extended Ricci rotation coefficient
$$\Upsilon^{\alpha}_{\;\; \mu\nu} \equiv h_I^{\;\alpha}h^I_{\;\mu ;\nu} +
h_i^{\;\alpha}h^I_{\;\mu}L^i_{I \, ,\nu} \qquad . \tag{3}$$ When $L^i_I$ is constant,
then the second term is zero and we have the usual definition, and also, in this case,
we have $\Upsilon^\alpha_{\;\;\mu\nu}= h_i^{\;\alpha} h^i_{\;\mu ; \nu}$. Henceforth we
will use the symbol $\Upsilon^{\alpha}_{\;\;\mu\nu}$ to mean the extended Ricci
rotation coefficient. One may easily verify that $\Upsilon^{\alpha}_{\;\;\mu\nu}$ is a
tensor and is a Lorentz scalar. We also have from this definition
$\Upsilon^I_{\;\;\mu\nu}=L^I_i \Upsilon^i_{\;\;\mu\nu}= L^I_i h^i_{\; \mu ; \nu} =
h^I_{\; \mu ;\nu} + L^I_i h^J_{\; \mu} L^i_{J \, , \nu}\;$.

\par \phantom{D} \par

{\bf  2. The Stroke Covariant Derivative.}
\par \phantom{D} \par

We now define a derivative which is covariant under more general coordinate
transformations on $x^\mu$ as well as local frame transformations on $x^I$. We will
call this extended covariant derivative the {\it stroke covariant derivative} will
denote it by use of a vertical stroke. An extended covariant derivative is a standard
device used in gauge theory and in the standard model (Ryder, 1996).  We anticipate
that our extended covariant derivative will be used to unify gravity with the other
forces. When acting on a contravariant vector, the stroke derivative is defined by
$$ \aligned V^\mu_{\;\; | \nu} & \equiv V^\mu_{\; ,\nu} + V^\beta h_i^{\;\mu}h^i_{\;\beta ,\nu}  \\
& \equiv  V^\mu_{\; ,\nu}+ V^\beta \biggl( h_I^{\,\,\mu} h^I_{\,\,\beta ,\nu}+  h^I_{\;
\beta} h_i^{\,\,\mu} L^i_{I,\nu}  \biggr) \quad . \endaligned \tag{4}
$$
As stated above, $x^i$ is inertial, $x^I$ is internal, and the field variables are
$L^i_I$ and $h^I_{\,\,\alpha}$.  The covariant derivative of the tetrad is $h^i_{\; \mu
;\nu}=h^i_{\; \mu ,\nu}- h^i_{\;\beta} \Gamma^{\beta}_{\mu\nu}$. Thus $h^i_{\;\mu ,
\nu} = h^i_{\; \mu ;\nu}+h^i_{\;\beta} \Gamma^\beta_{\mu\nu}$ and hence
$h_k^{\;\mu}h^k_{\;\beta ,\nu}= h_k^{\;\mu}h^k_{\;\beta ;\nu}+ \Gamma^\mu_{\beta\nu}$ .
Thus we have $h_I^{\,\,\mu} h^I_{\,\,\beta ,\nu}+ h^I_{\; \beta} h_j^{\,\,\mu}
L^j_{I,\nu}= \Gamma^\mu_{\beta\nu} + \Upsilon^\mu_{\;\;\beta \nu}$, and so the stroke
derivative may be written
$$V^\mu_{\;\; |\nu} = V^\mu_{\; ; \nu} + V^\beta \Upsilon^\mu_{\;\;\beta \nu} \tag{5} $$
where $\Upsilon^\mu_{\;\;\beta \nu}$ is the extended Ricci rotation coefficient defined
in (3).

Many investigators have used an alternative covariant derivative with connection given
by $L^\alpha_{\;\;\mu \nu}= h_i^{\;\alpha} h^i_{\;\mu , \nu}$ which is covariant under
all coordinate transformations $x^\mu \to x^{\bar{\mu}}$, but does not extend to local
Lorentz transformations.  In Weinberg (1972), the connection for $V^i$ is
$\gamma^i_{\;\; j k}$ which is not equal to our $L^I_j L^j_{J,K}$.   Kibble (1961)
introduces 24 fields $A^{ij}_{\;\;\; k}$ with $A^{ij}_{\;\;\; \mu} = - A^{ji}_{\;\;\;
\mu}$ through which a connection $\Gamma^\alpha_{\;\;\mu\nu}$ is defined.  This
connection is non-symmetric in its lower indices.  Hehl, et. al. (1976) use a
connection given by $\Gamma^k_{\; ij} = \bigl\{\phantom{l}^{\, k}_{ij} \;\bigr\} -
K_{ij}^{\;\;\; k}$, where $K_{ij}^{\;\;\; k}$, the non-Riemannian part of the
connection, is called the contortion.  Also, these authors do not use the same
Lagrangian as in our theory (usually they use $\int R\sqrt{-g}\, d^4 x$ ).  Our
connection is formed directly from the tetrad $h^I_{\;\mu}$ and $L^i_I$ which are
considered to be the fundamental fields.   Because of the extended Ricci rotation
coefficient, the stroke covariant derivative defined by (4) and (5) is covariant with
respect to a wider group of transformations than these other extended covariant
derivatives.

For covariant vectors one gets
$$ \aligned V_{\mu | \nu} &= V_{\mu ,\nu}
- V_{\beta} h^{\; \beta}_I h^I_{\; \mu ,\nu} - V_i
h^I_{\;\mu} L^i_{I ,\nu}  \quad \\
&=V_{\mu ; \nu} - V_{\beta} \Upsilon^\beta_{\;\;\mu \nu}   \endaligned \tag{6}
$$ where, again, $x^i$ is assumed to be inertial and the extended Ricci rotation
coefficient is used in the second line.  Using (5) and (6), one may verify the product
rule holds: $(U^\mu V_\nu)_{|\alpha} = U^\mu_{\; |\alpha}V_\nu + U^\mu V_{\nu |
\alpha}$.  It is also easy to see that $(U^\mu V_\mu)_{|\alpha} = (U^\mu
V_\mu)_{,\alpha}$ as would be expected.  Analogous formulas hold for tensors of higher
rank. For example,
$$\aligned V^\alpha_{\;\;\beta |\mu}&=V^\alpha_{\;\;\beta ,\mu}+ V^\gamma_{\;\;\beta}
h_I^{\,\,\alpha} h^I_{\,\,\gamma,\mu}+V^I_{\;\;\beta} h^{\,\,\alpha}_j L^j_{I,\mu} -
V^\alpha_{\;\;\gamma} h^{\,\,\gamma}_I h^I_{\,\,\beta ,\mu} - V^\alpha_{\;\;j}
h^I_{\;\beta} L^j_{I ,\mu}
\\  &= V^\alpha_{\;\; \beta ; \mu} + V^\gamma_{\;\; \beta} \Upsilon^\alpha_{\;\; \gamma \mu}
 - V^\alpha_{\;\;\gamma} \Upsilon^\gamma_{\;\;\beta \mu} \endaligned \tag{7}
$$
We use (4) to define
$$ V^I_{\; |\nu} \; \equiv \; h^I_{\;\mu}V^\mu_{\;|\nu}\; = \; V^I_{\; ,\nu}+
V^J L^I_j L^j_{J ,\nu} \quad , \tag{8}$$ and using (6) we have
$$ V_{I |\nu} \; \equiv \; h_I^{\;\mu}V_{\mu |\nu} \; = \; V_{I ,\nu}
- V_J L^J_j L^j_{I , \nu}  \tag{9}   $$

Using the formulas (4) - (9), one may take stroke covariant derivatives of quantities
which involve both Latin and Greek indices.   Thus $$V^I_{\;\; \alpha | \beta} =
V^I_{\;\; \alpha ,\beta}+ V^K_{\;\; \alpha} L^I_j L^j_{K , \beta} - V^I_{\;\; \gamma}
h^{\;\gamma}_K h^K_{\; \alpha , \beta} - V^I_{\;\; k} h^K_{\; \alpha} L^k_{K , \beta}
$$
If we apply this result to the field variable $h^I_{\;\alpha}$, noting that
$h^I_{\;\mu} h_{k}^{\;\mu} = L^I_k$, the result is
$$\aligned h^I_{\; \alpha |
\beta} &= h^I_{\; \alpha ,\beta}+ h^K_{\; \alpha} L^I_j L^j_{K , \beta} -
h^I_{\;\gamma} h^{\;\gamma}_K h^K_{\; \alpha , \beta} - L^I_k h^K_{\; \alpha} L^k_{K , \beta}  \\
&= h^I_{\; \alpha ,\beta}+ h^K_{\; \alpha} L^I_j L^j_{K , \beta} - h^I_{\;\alpha ,
\beta} - L^I_k h^K_{\; \alpha} L^k_{K , \beta} \\
&= 0 \endaligned \tag{10}
$$
It is an easy matter to verify that under general coordinate transformations,
$V^{\bar{\alpha}}_{\;\; |\nu}=x^{\bar{\alpha}}_{\,\, ,\mu}V^\mu_{\;\; | \nu}$ and also
under general Lorentz transformations that $V^{\hat{I}}_{\;\; | \alpha}= L^{\hat{I}}_J
V^J_{\;\; |\alpha}$.  Hence the stroke derivative of a vector or tensor is another
vector or tensor with a rank increased by one.

We also define $V^i_{\; | \nu} \equiv L^i_I V^I_{\; | \nu}$ and $V_{i |\nu} \equiv
L^I_i V_{I|\nu}$.  These definitions lead to
$$\qquad V^i_{\; |\nu} = V^i_{\; , \nu} \qquad \text{  and  } \qquad V_{i|\nu}= V_{i,\nu}\qquad , \tag{11}  $$
and we easily see that $$L^i_{I|\nu}= 0 \quad . $$

As a check on the consistency of the stroke covariant derivative and the fact that the
tetrad is stroke covariant constant we consider whether $\eta_{_{MN} \, |\nu}$ is zero
by direct calculation.  From (6) with use of the product rule, we have
$$ {\dsize \eta}_{_{MN} \, |\nu}
 =\eta_{_{MN },\nu} + \eta_{_{KN}}L^j_M L^K_{j , \nu} +
\eta_{_{MK}}L^j_N L^K_{j,\nu}  \quad . $$ Now ${\dsize \eta}_{_{MN},\nu}=0$.  Using
$L^J_j = \eta^{^{JK}} \eta_{jk} L^k_K\;$, we see that the second term reduces to the
negative of the third term:
$$\aligned  \eta_{_{KN}}L^j_M L^K_{j , \nu} =  -\eta_{_{KN}}L^j_{M ,\nu}L^K_j &=
-\eta_{_{KN}} \bigl(\eta^{ij} \eta_{_{MI}} L^I_i\bigr)_{ , \nu} \eta^{^{KL}}
\eta_{jk}L^k_L \\
&=-\eta_{_{KN}} \eta^{ij}\eta_{_{MI}}
\eta^{^{KL}} \eta_{jk}L^k_L L^I_{i , \nu} \\
&=-\delta^L_N \delta^i_k \eta_{_{MI}}
L^k_L L^I_{j ,\nu} \\
&=-\eta_{_{MI}} L^i_N L^I_{i ,\nu} \\
\endaligned $$
and hence  $$\eta_{_{MN} \, | \nu}=0 \qquad . $$

Let $\tilde{V}^\alpha$ be a vector density of weight $+1$ which may be constructed by
multiplying a vector $V^\alpha$ by $h= \sqrt{-g}$, the determinant of $h^i_{\,\,\mu}$.
Since $g_{\mu\nu ;\alpha}=0\,$, then $h_{\, ; \alpha}=0\,$.  It is also well known that
$\tilde{V}^\alpha_{\;\; ;\alpha} = \tilde{V}^\alpha_{\;\; , \alpha}$.  Also $h^i_{\,\,
\mu | \nu}=0\,$ implies that $\tilde{V}^\alpha_{\;\; |\alpha}= \bigl(h \,
V\bigr)^\alpha_{\;\; |\alpha}= h \, V^{\alpha}_{\;\; |\alpha}\,$, and hence one may
obtain the following rule for the stroke covariant divergence of vector density of
weight $+1$:
$$ \tilde{V}^\alpha_{\;\; |\alpha} =\tilde{V}^\alpha_{\;\; , \alpha}+
\tilde{V}^\beta \Upsilon^\alpha_{\;\; \beta \alpha} \quad . \tag{12}$$

{\bf Definition:}  The curvature vector (see Pandres, 1981, 1984) is given by
$$C_\mu \equiv \Upsilon^\alpha_{\;\;\mu\alpha}  \tag{13a} $$
The derivative of $h$ is given by $h_{\, ,\alpha}=h\,h_k^{\;\beta} h^k_{\; \beta
,\alpha}$. Since the extended Ricci rotation coefficient is used, this is an extension
of Pandres definition, but as its value is the same in the inertial coordinates, $x^i$,
no confusion will arise by using the same symbol, $C_\mu$. Using this and the
properties of covariant derivatives and the extended Ricci rotation coefficient one
finds that
$$\aligned C_\mu &= h_I^{\; \alpha} h^I_{\; \mu ; \alpha} + h^I_{\; \mu} L^i_{I ,i} \\
&= h_i^{\,\,\alpha}\bigl(h^i_{\,\, \mu ,\alpha}-h^i_{\,\, \alpha ,\mu} \bigr)
\\ &= h_I^{\;\alpha}\bigl(h^I_{\;\mu ,\alpha}-h^I_{\;\alpha , \mu}\bigr) + h^I_{\;\mu}
L^J_j(L^j_{I,J} - L^j_{J,I})
\endaligned
\tag{13b}$$   and
$$ \aligned C_i &= -h^{-1}\bigl(h\,h_i^{\;\alpha}\bigr)_{ ,\alpha} \\
C_I &= -h^{-1}\bigl(h\, h_I^{\;\alpha}\bigr)_{ ,\alpha} + L^i_{I,i} \\
C_I&= -H^{-1}\Bigl(H H_I^{\;\alpha}\Bigr)_{\,,\alpha} + \Lambda
\Bigl(\Lambda^{-1}\Lambda^i_I\Bigr)_{\,,i} \quad ,
\endaligned \tag{13c}
$$
where the last line, listed here for easy reference, will be explained in the next
section. It is easy to verify that $C_I$ transforms as a vector under {\it all
differentiable Lorentz transformations} on the Latin indices, i.e.
$C_{\bar{I}}=L^I_{\bar{I}} C_I$, provided $L^i_{\bar{I}}$ is differentiable. However,
for $C_\alpha $ to transform as a vector under changes of coordinates, $x^\alpha \to
x^{\bar{\alpha}}$, the transformation must be conservative, i.e.
$$C_{\bar{\alpha}}=x^\alpha_{\; ,\bar{\alpha}}C_\alpha \; \iff \; x^\nu_{\; ,
\bar{\alpha}}\biggl(x^{\bar{\alpha}}_{\; ,\mu ,\nu} - x^{\bar{\alpha}}_{\; , \nu , \mu}
\biggr) = 0 $$

\par \phantom{D} \par \phantom{D} \par

{\bf 3. Complex Lorentz transformations.  Complexification of the tetrad.}
\par\phantom{D} \par

We consider allowing the $h^I_{\;\mu}$ and $L^I_i$ to be complex.  We will denote
complex $h^I_{\;\mu}$ by $H^I_{\;\mu}$ and complex $L^I_i$ by $\Lambda^I_i$.   Note
that $h^i_{\;\mu}$ remains real and thus $g_{\mu\nu}$ remains real. When the Lorentz
group is extended to complex values, we will denote the transformation coefficients by
$\Lambda^{\bar{I}}_I$. There are two possible ways of extending (see Barut(1980)), one
in which $\eta_{\hat{I}\hat{J}}=\eta_{IJ}\Lambda^I_{\hat{I}} \Lambda^J_{\hat{J}}$, but
we extend the Lorentz group via the second possibility, i.e.,
$$\eta_{\hat{I}\hat{J}}=\eta_{IJ}\overline{\Lambda^I_{\hat{I}}} \Lambda^J_{\hat{J}} \quad , \tag{14}$$
where $\eta_{\hat{I}\hat{J}}=\eta_{IJ}=\text{diag}(-1,1,1,1)$ and a bar over a quantity
indicates its complex conjugate.  Since $\eta_{\hat{I}\hat{J}}$ is real then ${\dsize
\eta}_{\hat{I}\hat{J}}=\overline{\phantom{l}{\dsize\eta}_{\hat{I}\hat{J}}}= \eta_{IJ}
\Lambda^I_{\hat{I}} \overline{\Lambda^J_{\hat{J}}}$, and we see that
$\overline{\phantom{l}\Lambda^I_{\hat{I}}}$ is also a Lorentz transformation.

As before, we denote the inverse of $\Lambda^I_i$ as $\Lambda^i_I$ and convert between
the $x^i$ system and the $x^I$ system as usual, e.g. $V^I=V^i \Lambda^I_i$ and $V^i=V^I
\Lambda^i_I$. We also note that $\overline{\Lambda^I_i}$, the complex conjugate of
$\Lambda^I_i$ is also used to convert between the $x^i$ and $x^I$ system and the
inverse is the complex conjugate of $\Lambda^i_I$, i.e. $\overline{\Lambda^I_i}\,
\overline{\Lambda^i_J} =\delta^I_J$ and $\overline{\Lambda^I_j} \,
\overline{\Lambda^i_I}=\delta^i_j$. Let $V_I \equiv V_i \Lambda^i_I$, and $V^I\equiv
V^i \Lambda^I_i$. Then $\overline{V_I} = V_i \overline{\Lambda^i_I}$ and
$\overline{V^I} = V^i \overline{\Lambda^I_i}$.  Similar rules apply for tensors. For
the complex tetrad $H^I_{\;\alpha}$, one finds that $H^I_{\;\alpha} = \Lambda^I_i
h^i_{\;\alpha} $ has inverse $H_I^{\;\alpha}=h_i^{\;\alpha}\Lambda^i_I$ and that
$\overline{H^I_{\;\alpha}}= \overline{\Lambda^I_i} h^i_{\;\alpha} $ has inverse
$\overline{H_I^{\;\alpha}}=h_i^{\;\alpha}\overline{\Lambda^i_I}$.  Note that general
complexification leads to the condition that $g_{\nu \mu}=\overline{g_{\mu\nu}}$, but
because the $x^i$ and $x^\mu$ spaces remain real in our construction, $g_{\mu\nu}$
remains real and hence remains symmetric.

The determinant of $H^I_{\;\alpha}$ will be denoted by $H$.  We also define $\Lambda
\equiv \text{det}(\Lambda^i_I)$ and thus $h= H\Lambda$.  When inversions are excluded,
and $x^I$ is real, then $\Lambda=1$, and thus $h= H $; when the $\Lambda^i_I$ is
non-real, then $\Lambda=e^{i\theta}$ and hence $h=He^{i\theta}$, where generally
$\theta$ is a function of position $\theta(x)$. These comments explain the last line of
equations (13c).

When raising or lowering indices, complex conjugation must be used.  One finds that
$V^I={\dsize \eta}^{IJ} \overline{V_J}$ and $V_I={\dsize \eta}_{IJ} \overline{V^J}$.
Thus $V^IV_I= {\dsize \eta}^{IJ} \overline{V_J}V_I= \overline{V^I}\,
\overline{V^{\phantom{i}}_I}$. One also finds that $H^I_{\;\alpha} = {\dsize \eta}^{IJ}
g_{\alpha \beta} \overline{H_J^{\;\,\beta}} $. The definition for the extended Ricci
rotation coefficient is $\Upsilon^\alpha_{\;\;\mu\nu} = H_I^{\;\alpha}H^I_{\;\mu
;\nu}+H_i^{\;\alpha}H^I_{\;\mu}\Lambda^i_{I ,\nu} $, and the curvature vector, $C_I$,
is given by (13c).  These quantities are invariant under local Lorentz transformations
and conservative transformations on Greek indices. The stroke derivative is invariant
under local complex frame transformations.

There are a couple of reasons for extending the group of transformations to include the
complex Lorentz transformations.  It is well known (Barut, 1980) that the complex
Lorentz group which satisfies (14) contains $SU(3)$ as a proper subgroup and that
complex quantities are required for $SU(3)$. The complex Lorentz group, $\varLambda$,
has 16 parameters. Also, the inclusion of spinors and the spinor connection imply that
complex Lorentz transformations should be included.

 \par \phantom{D} \par \phantom{D} \par \phantom{D}

{\bf 4. The Field Lagrangian.}
\par \phantom{D} \par

We know that in general relativity we have the property that for a vector density of
weight $+1$, $\tilde{V}^\alpha_{\;\; ;\alpha}\equiv \tilde{V}^\alpha_{\;\; ,\alpha}$.
Thus an appropriate measure of the new geometry should be $$\tilde{V}^\alpha_{\;\;
|\alpha}-\tilde{V}^\alpha_{\;\; ,\alpha} = \tilde{V}^\alpha C_\alpha \quad .\tag{15}$$
The line of reasoning that leads to this conclusion is as follows. In flat space with a
continuously twice-differentiable vector $V^\alpha$, we have $V^{\alpha}_{\;\; ,\mu
,\nu}-V^{\alpha}_{\;\; ,\nu ,\mu}=0$.  Upon replacing the ordinary derivatives by
covariant derivative we use  $V^{\alpha}_{\;\; ; \mu ; \nu }- V^\alpha_{\;\; ; \nu ;
\mu}= -V^{\beta}R^\alpha_{\;\;\beta \mu\nu}$ to measure the non-flatness of the
corresponding Riemannian geometry. The curvature tensor, $R^\alpha_{\;\;\beta \mu\nu}$,
transforms as a tensor under diffeomorphisms.  In a similar way, a space is
conservatively flat with respect to the conservation group when $\tilde{V}^\alpha_{\;\;
; \alpha}-\tilde{V}^\alpha_{\;\; ,\alpha} = 0$ and hence, after replacing the covariant
derivative with the stroke covariant derivative,  the non-flatness of the conservation
geometry is measured by (15).  The quantity $C_\mu$ transforms as a vector under
conservative transformations and $C_I$ transforms as a vector under all differentiable
Lorentz transformations.  We note that there exists a conservative transformation
between $x^\alpha$ and $x^{\hat{\alpha}}$ such that $g_{\hat{\alpha}\hat{\beta}}=
\text{diag} (-1,1,1,1)$ if and only if $C_\mu=0$ (Pandres, 1981).

A suitable field Lagrangian will be a scalar which is constructed from $C_\mu$. Thus a
suitable field Lagrangian is given by
$$\Cal{L}=\int C^\alpha C_\alpha \, h \; d^4 x \tag{16}  $$ where $h = \sqrt{-g}$ is the
determinant of the tetrad $h^i_{\;\alpha}$. We also have $\Cal{L}=\int C^iC_i \, h \;
d^4x $ and $\Cal{L}=\int C^IC_I \, H\Lambda\; d^4x$.   The Riemann tensor is given by
$R^\alpha_{\;\;\beta \mu\nu}= h_i^{\;\alpha}(h^i_{\;\beta ;\mu ;\nu}-h^i_{\;\beta ;\nu
;\mu})$.  Using (3) one finds that the Riemann tensor, the Ricci tensor and the Ricci
scalar are given by
$$\aligned R^\alpha_{\;\;\beta\mu\nu}&= \Upsilon^\alpha_{\;\;\beta \mu
;\nu}-\Upsilon^\alpha_{\;\;\beta\nu ;\mu}+ \Upsilon_{\;\;\sigma \nu}^\alpha
\Upsilon^\sigma_{\;\;\beta \mu} -\Upsilon^\alpha_{\;\;\sigma \mu}
\Upsilon^\sigma_{\;\;\beta\nu} + h_I^{\;\alpha}h^i_{\;\beta}
(\Lambda^I_{i,\mu,\nu}-\Lambda^I_{i,\nu
,\mu}) \\
R_{\mu\nu} \; &= C_{\mu ; \nu} - \Upsilon^\alpha_{\;\;\mu\nu ;\alpha}+
\Upsilon^\alpha_{\;\;\sigma \nu} \Upsilon^\sigma_{\;\;\mu\alpha} -
\Upsilon^\alpha_{\;\;\mu\nu} C_\alpha +
h_I^{\;\alpha}h^i_{\;\mu}(\Lambda^I_{i,\alpha,\nu}-\Lambda^I_{i,\nu\alpha}) \\
R \;\; &= 2C^\alpha_{\;\; ;\alpha}+C^\alpha C_\alpha - \Upsilon^{\alpha\beta\nu}
\Upsilon_{\alpha \nu \beta} + \eta^{ij}h^{\;\nu}_j h_I^{\;\alpha}
(\Lambda^I_{i,\alpha,\nu}-\Lambda^I_{i,\nu,\alpha})\quad .
\endaligned \tag{17}$$ Thus one finds that (see Green and Pandres, 2003)
$$C^\alpha C_\alpha = R + \Upsilon^{\alpha \beta \nu} \Upsilon_{\alpha \nu \beta}-2C^\alpha_{\;
;\alpha} - \eta^{ij}h_j^{\;\nu}h_I^{\;\alpha}\bigl(\Lambda^I_{i ,\alpha
,\nu}-\Lambda^I_{i,\nu ,\alpha}\bigr) \quad . \tag{18} $$ The additional terms are
suggestive of non-gravitational interactions.

Setting $\delta \Cal{L} \, = 0$ leads to field equations.  The fields that will be
varied are $H^I_{\;\alpha}$ and $\Lambda^i_I$. The requirement that $\eta_{IJ}=
\text{diag}(-1,1,1,1)$ and the requirement that $h^i_{\;\mu}=H^I_{\;\mu}\Lambda^i_I$ be
real will not be imposed at the outset by using Lagrange multipliers.  Nevertheless the
resulting field equations will have solutions with these properties and hence these
constraints do not affect the variational problem. Now, $\delta (C^I C_I H\Lambda) =
(C^I C_I) \,\Lambda\, \delta H + (C^IC_I) \, H\, \delta \Lambda + 2 H\Lambda C^I
\,\delta C_I$. Thus, from the formulas $\delta H = (-H \, H^K_{\;\;\nu} )\, {\delta
H}^{\;\;\nu}_K$ and $\delta \Lambda = \Lambda \Lambda^J_j \, \delta \Lambda^j_J$ and
using (13c) we easily find that $\delta C_I = -H^{-1}H^K_{\;\nu} (H
H_I^{\;\alpha})_{,\alpha}\,\delta H^{\;\nu}_K - H^{-1}\Bigl(\delta(H H^{\;\alpha}_I)
\Bigr)_{ ,\alpha} + \Lambda \Lambda^J_j \Bigl(\Lambda^{-1}\Lambda^i_I\Bigr)_{,i} \,
\delta \Lambda^j_J + \Lambda \Bigl( \, \delta(\Lambda^{-1}\Lambda^i_I)\Bigr)_{, i} \;$.
When these results are used and an integration by parts is performed, one obtains
$$ \aligned \delta (C^IC_I H\Lambda)  = \;&
-2\Lambda C^I\bigl(H H_I^{\;\alpha}\bigr)_{,\alpha} H^K_{\;\nu} \,\delta H_K^{\;\nu} +
2\bigl(\Lambda C^I\bigr)_{,\alpha} \delta(H H_I^{\;\alpha}) \\
& + 2H\Lambda^2C^I\bigl(\Lambda^{-1}\Lambda^i_I\bigr)_{,i}\Lambda^J_j\, \delta
\Lambda^j_J -
2\bigl(H\Lambda^2C^I H_i^{\;\alpha}\bigr)_{,\alpha} \delta(\Lambda^{-1}\Lambda^i_I) \\
& - H\Lambda C^IC_I H^K_{\;\nu} \, \delta H_K^{\;\nu} + H\Lambda C^I C_I \Lambda^J_j \,
\delta \Lambda^j_J \quad ,
\endaligned $$
where the boundary terms have been discarded since $\delta (H^\nu_K)=0$ and $\delta
(\Lambda^i_I)=0$ on the boundary.  After straightforward use of the product rule and
chain rule, one obtains
$$ \aligned \delta(C^I C_I H\Lambda) =&  2H\Lambda \biggl(\tfrac12 \,C^IC_I
H^K_{\;\nu}-C^I\Lambda^i_{I,i} H^K_{\;\nu} +\Lambda^J_j\Lambda^j_{J,\nu}C^K
+C^K_{,\nu}- C^I_{,I} H^K_{\;\nu}\biggr)\delta H_K^{\;\nu}
\\ &  + 2H\Lambda\biggl(  C^I \Lambda \Bigl(\Lambda^{-1}\Lambda^i_I\Bigr)_{,i}
\Lambda^J_j -2\Lambda^K_k \Lambda^k_{K,j}C^J +
2\Lambda^K_k\Lambda^k_{K,I}C^I\Lambda^J_j
\\ &-C^J_{,j}+C^I_{,I}\Lambda^J_j +C_jC^J-
\tfrac12 \, C^IC_I \Lambda^J_j \biggr)\delta \Lambda^j_J
\endaligned \tag{19} $$
Since $h=H\Lambda$ must be nonzero and since $\delta H^\nu_K$ is arbitrary in the
region of integration, $\delta \Cal{L}=0$ implies that the expression in the first
parenthesis in (19) must be zero.  Multiplying this expression by $H^{\;\nu}_L$ one
obtains
$$ \tfrac12 \,C^IC_I \delta^K_L - C^I \Lambda^i_{I,i} \delta^K_L  + \Lambda^J_j \Lambda^j_{J,L}C^K +
C^K_{,L} - C^I_{,I} \delta^K_L = 0  \quad . \tag{20}$$ The trace of this equation
implies that $$ 2C^IC_I - 4C^I \Lambda^i_{I,i}+C^I \Lambda^J_j \Lambda^j_{J,I} -
3C^I_{,I} = 0 \tag{21}$$

Similarly the expression in the second parenthesis of (19) must be zero also.
Multiplying this expression by $\Lambda^j_L$ one finds that
$$\delta^J_L C^I \Lambda^i_{I,i}+\delta^J_L C^I \Lambda^K_k \Lambda^k_{K,I} -
2\Lambda^K_k \Lambda^k_{K,L} C^J- C^J_{,L}+\delta^J_L C^I_{,I}+C^J C_L -
\tfrac12\delta^J_L C^IC_I = 0 \quad . \tag{22}$$ The trace of this equation yields
$$C^I C_I - 4C^I \Lambda^i_{I,i}-2C^I \Lambda^J_j \Lambda^j_{J,I} - 3C^I_{,I}=0 \tag{23}$$
and hence subtracting (23) from (21) gives $C^I \Lambda^J_j \Lambda^j_{J,I} =
-\frac13C^I C_I$. Also multiplying (21) by 2 and adding to (23) yields $C^I
\Lambda^i_{I,i}=\frac5{12}C^IC_I-\frac34 C^I_{,I}$.  After inserting these formulae
into (20) and (22), one obtains
$$ \frac1{12} \delta^K_L C^I C_I-\frac14 \delta^K_L C^I_{,I} +\Lambda^J_j \Lambda^j_{J,L} C^K +
C^K_{,L} = 0$$ and
$$-\frac5{12} \delta^K_L C^I C_I + \frac14\delta^K_L C^I_{,I} -2\Lambda^J_j \Lambda^j_{J,L}C^K
-C^K_{,L} + C^KC_L = 0 \quad . $$ The sum of these two equations yields
$$C^K C_L - \Lambda^J_j \Lambda^j_{J,L} C^K = \frac13 \delta^K_L C^I C_I \quad . $$
Now since $\Lambda$ is the determinant of a complex lorentz transformation,
$\Lambda=e^{i\theta}$ and thus $\Lambda^J_j \Lambda^j_{J,K}=
\frac{\Lambda_{,K}}{\Lambda} = i\,\theta_{,K}$. Thus
$$ C^K C_L - i\,\theta_{,L}C^K = \frac13 \delta^K_L C^I C_I \quad . \tag{24}$$

Now multiply equation (24) by $C_K$ and sum over $K$.  Assume that $C^K C_K\neq 0$.
Then this implies that $C_{L}-i\theta_{,L} = \frac13 C_L$ and hence $C_L=\frac32 i
\theta_{,L}$. Substituting this into (24) leads to $C^K C_K=0$ which contradicts our
assumption.  Thus we see that our field equations imply that $C^K C_K=0$.

From (24), we now see that $C^K(C_L-i\theta_{,L})=0$.  Now assume that $C^K\neq 0$ and
substitute $C_L=i\theta_{,L}$ into (20).  Then when $K\neq L$, this implies that
$i\theta^{,K}_{\;\; ,L}= \theta^{,K} \theta_{,L}$.  But since $\theta$ is real then
$\theta^{,K}\theta_{\;\; ,L}=0$ when $K\neq L$.  Thus at most one of the $\theta_{,L}$
is nonzero, but then $C^K C_K=0$ would imply that all are zero, contradicting the
assumption that $C^K\neq 0$.

Hence the field equations imply that $C_I=0$ and since $C_\alpha = C_I h^I_{\;\alpha}$,
we have $$ C_\alpha = 0 \quad . \tag{25}$$

There are several examples of solutions to the field equations (25). The first example
is given by $h^i_{\;\mu} = \delta^i_\mu + \delta^i_0\delta^2_\mu x^1$, where $x^1$ is a
Greek coordinate value, (see Pandres, 1981), and results in a Ricci scalar value of
$R=\frac12$.   This is equivalent to the pair: $h^I_{\;\mu} = \delta^I_\mu +
\delta^I_0\delta^2_\mu x^1$ and $L^i_I=\delta^i_I$. A second example is given by
$$h_i^{\;\mu} = \delta^\mu_0\delta^0_i +\delta^\mu_3 \delta^3_i +
(\delta^\mu_1 \delta^1_i + \delta^\mu_2\delta^2_i)\cos
x^3+(\delta^\mu_2\delta^1_i-\delta^\mu_1\delta^2_i)\sin x^3 \quad , \tag{26}
$$ where $x^3$ is a Greek coordinate.  For (26), $g_{\mu\nu}= \text{diag}(-1,1,1,1)$
and hence $R^\alpha_{\;\;\beta\mu\nu}=0$, but $\Upsilon^\alpha_{\;\;\mu\nu}\neq 0$.  A
third example is a spherically symmetric solution of the field equations.  Let $f(r)$
be a positive differentiable function of $r=\sqrt{(x^1)^2+(x^2)^2+(x^3)^2}$. Then the
tetrad given by
$$h^i_{\;\; \mu} = \delta^i_0 \delta^0_\mu \sqrt{f(r)} + \frac1{\root 4 \of{f(r)}} (\delta^i_1
\delta^1_\mu + \delta^i_2 \delta^2_\mu + \delta^i_3 \delta^3_\mu ) \tag{27} $$ yields
$C_\mu=0$ and hence is a solution of the field equations.  The metric, in line element
form,  is given by
$$ ds^2 = -f(r)dt^2 + \frac1{\sqrt{f(r)}} dr^2 + \frac{r^2}{\sqrt{f(r)}} d\theta^2 +
\frac{r^2\sin^2\theta}{\sqrt{f(r)}} d\phi^2 \quad , \tag{28} $$ and both
$R^\alpha_{\;\;\beta\mu\nu}$ and $\Upsilon^\alpha_{\;\;\mu\nu}$ are nonzero.

Using the Einstein tensor $G_{\mu\nu}=R_{\mu\nu}-\frac12\,g_{\mu\nu}R$, the field
equations (25) and symmetrizing (so that $G_{\mu\nu}=G_{\nu\mu}$) we find that
$$\aligned G_{\mu\nu} &= -\tfrac12\bigl(\Upsilon^\alpha_{\;\;\mu \nu ;\alpha} +
\Upsilon^\alpha_{\;\;\nu \mu ;\alpha}  \bigr)    +\tfrac12
\bigl(\Upsilon^\alpha_{\;\;\sigma \nu} \Upsilon^\sigma_{\;\; \mu
\alpha}+\Upsilon^\alpha_{\;\;\sigma \mu} \Upsilon^\sigma_{\;\; \nu \alpha} \bigr)  +
\tfrac12\, g_{\mu\nu}\Upsilon^{\alpha \beta
\sigma} \Upsilon_{\alpha \sigma \beta}\quad \\
& \qquad + \tfrac12\Bigl(
h_I^{\;\alpha}h^i_{\;\mu}(\Lambda^I_{i,\alpha,\nu}-\Lambda^I_{i,\nu,\alpha}) +
h_I^{\;\alpha}h^i_{\;\nu}(\Lambda^I_{i,\alpha,\mu}-\Lambda^I_{i,\mu,\alpha}) \Bigr) \\
& \qquad - \tfrac12\,g_{\mu\nu}\eta^{ij}h_j^{\;\sigma}
h_I^{\;\alpha}(\Lambda^I_{i,\alpha,\sigma}-\Lambda^I_{i,\sigma,\alpha}) \quad .
\endaligned \tag{29}
$$ These terms on the right suggest that, when interpreted in Riemannian geometry, this
new geometry may automatically produce an appropriate stress energy tensor.

\par \phantom{D} \par

{\bf 5. Inclusion of spinors.  The spin connection.}

\par\phantom{D} \par

The fundamental constant spin tensors, $\sigma^{i\dot{a} }_{\;\;\;\; b}$, are given as
follows (Bade and Jehle, 1953; Clarke and de Felice, 1992).
$$\matrix
\sigma^{0\dot{a}}_{\;\;\;\;b}&=\frac1{\sqrt{2}}\bmatrix 0 & 1\\-1 & 0\endbmatrix \quad
,& \quad
\sigma^{1\dot{a}}_{\;\;\;\;b}&=\frac1{\sqrt{2}}\bmatrix -1 & 0 \\ 0 & 1 \endbmatrix \\
 & \phantom{D}& & \\
\sigma^{2\dot{a}}_{\;\;\;\;b}&=\frac1{\sqrt{2}}\bmatrix i & 0 \\0 & i\endbmatrix \quad
,& \quad \sigma^{3\dot{a}}_{\;\;\;\;b}&=\frac1{\sqrt{2}}\bmatrix 0 & 1 \\ 1 & 0
\endbmatrix
\endmatrix \tag{30}
$$  We typically will use Latin indices $a$ through $f$ for spin indices (first index refers
to the row and the second index refers to the column), and $\sigma^{i\dot{a}b}$ is
defined by $\sigma^{i\dot{a}b}= -\sigma^{i\dot{c}}_{\;\;\;\;d} \Cal E^{db}$ and also
$\sigma^{i}_{\;\;\dot{a}b}= -\Cal E_{\dot{a}\dot{c}} \sigma^{i\dot{c}}_{\;\;\;\; b} $,
where the spin metric is given by
$$\Cal E^{ab}=\Cal E_{ab}=\Cal E^{\dot{a}\dot{b}}=\Cal E_{\dot{a}\dot{b}}=\bmatrix 0 &
1 \\ -1 & 0 \endbmatrix \quad . \tag{31}$$  Note that $\pmb{\Cal E}$ is antisymmetric.
When we use matrix multiplication to aid in the computation process, we lower indices
via a sum on adjacent indices with the matrix for $\Cal E$ afterward (or sum on
adjacent indices with the matrix for $-\Cal E$ before the spinor).   Similarly, when we
use matrix multiplication in raising indices, we sum on adjacent indices with the
matrix for $\Cal E$ before (or sum on adjacent indices with the matrix for $-\Cal E$
afterward).  Basically, when raising or lowering spinor indices, the summed indices
should be adjacent and the sign is $+$ for $\searrow$ and $-$ for $\nearrow$.  Useful
relations between the $\sigma^i$'s are: $\;\; \sigma^i_{\;\;\dot{a}b}\sigma_j^{\;\;
\dot{a}b}=-\delta^i_j \; \;$  ,  $\;\;
\sigma^i_{\;\;\dot{a}b}\sigma_i^{\;\;\dot{c}d}=-\delta^{\dot{c}}_{\dot{a}}
\delta^d_b\;\;$ and $\;\; \sigma^{i a}_{\;\;\;\; \dot{c}}\sigma^{j\dot{c}}_{\;\;\;\;
b}+ \sigma^{j a}_{\;\;\;\; \dot{c}}\sigma^{i\dot{c}}_{\;\;\;\; b}=\eta^{ij}\delta^a_b$
. When the meaning is clear we will suppress the spinor indices, for example
$\bold{\sigma}^i$ and $\bold{\Cal E}$.

Generally, for second rank spinors (with $2\times 2$ matrix representation) such as
$M^a_{\;\;c}$, we have $\Cal E_{ab} M^a_{\;\;c} M^b_{\;\;d} = det(M) \Cal E_{cd}$.
Thus, if $A^a_{\;\;c}$ has determinant $+1$, then $\Cal E_{ab} A^a_{\;\;c}
A^b_{\;\;d}=\Cal E_{cd}$, i.e. the metric is preserved. We will call these
$A^a_{\;\;b}$ spin transformations and they are elements of $SL(2,\Bbb C)$.  The real
Lorentz group is a 6 parameter group as is $SL(2,\Bbb C)$. As is usual in the tetrad
formalism, the fundamental spin tensors are kept constant by coordinating a spin
transformation, $A^a_{\hat{b}}\; \in SL(2,\Bbb C)$, with the Lorentz transformation
$L^i_{\hat{j}}$.  Since $L^I_i$ are field variables, these induce field variables
$A^A_a$.  This is because we keep $\sigma^{I\dot{A}}_{\;\;\;\;B}$ identical to
$\sigma^{i\dot{a}}_{\;\;\;\; b}$ by coordinating $A^A_a$ with the field variables
$L^I_i$.  As noted above, we only allow constant (global) Lorentz transformations,
$L^i_{\bar{j}}$, on the $x^i$ (inertial) space and hence we only allow constant
$A^a_{\hat{b}}$ on the corresponding inertial spinor space.  On the internal space,
$x^I$ and its corresponding spinor space, we allow nonconstant (local) Lorentz
transformations and nonconstant (local) spin transformations.

Now there is a 1-1 mapping from vectors $V^i$ to rank 2 spinors $V^{\dot{a}b}$ via
(31). Specifically $V^{\dot{a}b}=\sigma_i^{\;\dot{a}b}V^i$ which via the relation
$\sigma^i_{\;\; \dot{a} b} \sigma_j^{\;\; \dot{a} b} = - \delta^i_j$  implies
$V^i=-\sigma^i_{\;\dot{a}b}V^{\dot{a}b} $. Since there is coordination between the
field variables $L^i_I$ and the induced variables $A^A_a$, we also have
$V^I=-\sigma^I_{\;\dot{A}B} V^{\dot{A}B}$ and $V^{\dot{A}B}=\sigma_I^{\;\dot{A}B}V^I$.
Now, because of the constancy of the $\sigma\text{'s}\;$, $\; \sigma^i_{\;\dot{a}b
,\nu}=0$ and $\sigma^I_{\;\dot{A}B,\nu}=0$. From $\sigma^i_{\;\dot{a}b}=
\sigma^I_{\;\dot{A}B} L^i_I A^{\dot{A}}_{\dot{a}}A^B_b$, one finds that
$\sigma^I_{\;\dot{A}B} \Bigl(L^i_I A^{\dot{A}}_{\dot{a}} A^B_b \Bigr)_{,\nu} = 0$. Thus
$\sigma^I_{\;\dot{A}B} L^i_{I,\nu} =  - \sigma^I_{\;\dot{C}B}L^i_I
A^{\dot{a}}_{\dot{A}} A^{\dot{C}}_{\dot{a},\nu} - \sigma^I_{\;\dot{A} C} L^i_I A^a_A
A^C_{a,\nu}$.  Substituting this into the equation $V^I_{\;|\nu}=
\bigl(-\sigma^I_{\;\;\dot{A}B} V^{\dot{A}B}\bigr)_{\;|\nu}=
-\sigma^I_{\;\dot{A}B}V^{\dot{A}B}_{\phantom{AA},\nu}-\sigma^J_{\;\dot{A}B}V^{\dot{A}B}L^I_jL^j_{J,\nu}$,
we arrive at the spin form of the stroke covariant derivative of $V^I$,
$$V^I_{|\nu} = -\sigma^I_{\dot{A}B} \biggl( V^{\dot{A}B}_{\quad ,\nu} -
V^{\dot{C} B} A^{\dot{a}}_{\dot{C}} A^{\dot{A}}_{\dot{a} \, ,\nu} -  V^{\dot{A} C}
A^a_C A^B_{a \, ,\nu} \biggr) \quad . \tag{32}$$ Let $a_\mu$ be an arbitrary real
vector. One notices that, as in the usual spinor connection, that we may take the
replacement $A^a_B A^C_{a,\nu} \to A^a_B A^C_{a,\nu} + i\delta^C_B a_\mu $ which has no
effect on (32). This corresponds to the classical gauge transformation (see Bade and
Jehle). Thus a consistent definition for the stroke derivative of a spinor is given by
$$\Psi^A_{\;\; |\nu} = \Psi^A_{\;\; ,\nu} - \Psi^B \bigl(A^a_B A^A_{a,\nu} + i\delta^A_B a_\nu \bigr) \quad \tag{32a}$$
and
$$\Psi^{\dot{A}}_{\;\; |\nu} = \Psi^{\dot{A}}_{\;\; ,\nu} - \Psi^{\dot{B}} \bigl(
A^{\dot{a}}_{\dot{B}} A^{\dot{A}}_{\dot{a},\nu} -i\delta^{\dot{A}}_{\dot{B}} a_\nu
\bigr) \quad . \tag{32b}$$ These definitions imply that $\sigma^I_{\dot{A}B|\nu}=0  $.

We now consider the extension under parity from the 2-spinor to the 4-spinor.  The
indices for a 4-spinor will run from 1 to 4 with indices (1,2) corresponding to {\bf
dotted} 2-spinor indices and indices (3,4) corresponding to {\bf undotted} 2-spinor
indices. Let the $n\times n$ zero matrix be denoted by $0_n$ .  Let the matrices for
$\sigma^{i \dot{a}}_{\;\;\;\; b}$ be briefly denoted by $ \pmb{\sigma}^i \,$, then (in
the chiral form) the Dirac matrices, $\gamma^{i a}_{\;\;\;\;b}$ are given by
$$ \gamma^i \equiv \sqrt{2} \bmatrix 0_{2_{\phantom{D}}} & \pmb{\sigma}^i \\
\overline{\pmb{\sigma}^{i^{\phantom{i}}}}  & 0_2
\endbmatrix  \quad , \tag{33} $$ where the $\overline{\pmb{\sigma}^{i^{\phantom{i}}}}$ denotes
the complex conjugate (i.e. is $\sigma^{i a}_{\;\;\;\;\dot{b}}$). One finds that
$$\gamma_{i\;\; c}^{\; a} \gamma_{j\;\; b}^{\; c} + \gamma_{j\;\; c}^{\; a}
\gamma_{i\;\; b}^{\; c} = 2 \eta_{ij} \delta^a_b \quad , \tag{34a}$$ or in matrix
notation,
$$\gamma_i\gamma_j+\gamma_j\gamma_i = 2 \eta_{ij} I_4 \quad , \tag{34b}$$ where $I_4$
represents the $4\times 4$ identity matrix.  When the signature of the metric is
diag(-1,1,1,1), the Klein-Gordon equation is $\bigl(\partial^i\partial_i +
m^2\bigr)\Psi^b=0$ and the Dirac equation is given by $(\gamma^{i a}_{\;\;\;\; b}\, p_i
+ \, m \, \delta^a_b)\Psi^b = 0$.   In inertial coordinates, the Dirac equation is
$\bigl(i\gamma^{k a}_{\;\;\;\; b} \partial_k + m \delta^a_b \bigr)\Psi^b=0$ and upon
multiplying on the left by the operator $i\bold{\gamma}^j\partial_j$, one finds that
the Dirac equation implies the Klein-Gordon equation.

The metric tensor for 4-dimensional spinors is given by
$$\Cal E_{ab} \equiv  \delta_a^1\delta^2_b - \delta_a^2\delta^1_b + \delta_a^3\delta^4_b -
\delta_a^4\delta^3_b  \tag{35a}$$ and $\Cal E^{ab}=\Cal E_{ab}$.  Using (31) we have
the matrix form
$$ \Cal E_4  \equiv \bmatrix \Cal E & 0_2 \\ 0_2 & \Cal E \endbmatrix \quad . \tag{35b} $$
Suppose that $M^a_{\;\;c}$ has either of the following special forms:
$$M^a_{\;\;c}= \bmatrix 0_2 & A_2 \\
B_2 & 0_2 \endbmatrix \quad \text{ or } \quad M^a_{\;\;c}=\bmatrix A_2 & 0_2 \\ 0_2 &
B_2 \endbmatrix $$ where $A_2$ and $B_2$ are $2\times 2$ matrices with $det(A)=det(B)$.
Using (35) , we see that $  \Cal E_{ab} M^a_{\;\; c} M^b_{\;\; d} = det(A) \Cal E_{cd}
$.  Hence we define spinor transformations for 4-spinors by
$$\Bbb A^a_{\hat{b}} \equiv  \bmatrix \overline{A} & 0_2 \\ 0_2 & A \endbmatrix \quad , \tag{36}$$
where $\overline{A}$ is the complex conjugate of $A$ and both are elements of
$SL(2,\Bbb C)$.  We also see that there is a mapping between vectors and second rank
4-spinors given by $V^a_{\;\;\;b}=\frac12 \, V_i \gamma^{i\, a}_{\;\;\;\;b}$ with
inverse
 mapping given by $V^i=\frac12 \gamma^{i\, b}_{\;\;\;\;a} V^a_{\;\;\; b}$. As with
2-spinors, there is coordination between Lorentz transformations on the Latin indices
and spin transformations so that the $\Bbb \gamma^i$ remain constant. Similarly, when
the field variables $L^i_I$ are given, we require that $\gamma^{IA}_{\;\;\;\; B}$
remains unchanged and hence we see that this induces the values of $\Bbb A^A_a$. The
correspondence is exactly one-to-two, with $\Bbb A^A_a$ determined up to a sign. This
implies that $\bigl(\gamma^{I A}_{\;\;\;\;B} \Bbb A^a_A \Bbb A^B_b L^i_I \bigr)_{,\nu}
= 0 $ and thus $\gamma^{I A}_{\;\;\;\;B} \Bigl(L^i_I \Bbb A^a_A \Bbb A^B_b
\Bigr)_{,\nu} = 0$. From this we derive that $\gamma^{J A}_{\;\;\;\;B} L^I_i
L^i_{J,\nu} = -\gamma^{IC}_{\;\;\;\;B} \Bbb A^A_a \Bbb A^a_{C,\nu} +
\gamma^{IA}_{\;\;\;\; C} \Bbb A^C_a \Bbb A^a_{B,\nu}$.  Thus
$$ V^I_{\,|\nu} = \biggl(\frac12 \gamma^{IA}_{\;\;\;\;B} V^B_{\;\;A}\biggr)_{|\nu}
=\frac12\gamma^{IA}_{\;\;\;\;B}\biggl(V^B_{\;\;A,\nu}-V^B_{\;\;C}\Bbb A^C_a \Bbb
A^a_{A,\nu} + V^C_{\;\;A}\Bbb A^B_a \Bbb A^a_{C,\nu}\biggr) \quad . \tag{37}
$$
We note that, for arbitrary vector $a_\nu$,  the replacement $\Bbb A^B_a \Bbb
A^a_{C,\nu} \to \Bbb A^B_a \Bbb A^a_{C,\nu} + i\delta^B_C a_\nu$ has no effect on (37).
Thus we define the stroke derivatives of 4-spinors by
$$ \Psi^B_{\, |\nu} \equiv \Psi^B_{\, ,\nu} + \Psi^C \biggl(\Bbb A^B_a \Bbb A^a_{C,\nu} +
i \delta^B_C a_{\nu} \biggr) \; = \; \biggl(\partial_\nu + ia_\nu \biggr)\Psi^B +
\Psi^C \Bbb A^B_a \Bbb A^a_{C,\nu}   \tag{38}
$$
and
$$ \Psi_{A |\nu} \equiv \Psi_{A,\nu} - \Psi_C \biggl(\Bbb A^C_a \Bbb A^a_{A,\nu} +
i \delta^C_A a_{\nu} \biggr) \; = \; \biggl(\partial_\nu - ia_\nu \biggr)\Psi_A -
\Psi_C \Bbb A^C_a \Bbb A^a_{A,\nu} \quad . \tag{39}
$$
The definition for the stroke derivative of a 4-spinor implies that
$\gamma^{IA}_{\;\;\;\;B|\nu}=0$.


\par \phantom{D} \par

{\bf 6.  Concluding Remarks.}
\par\phantom{D} \par

We have established invertible transformations which convert between the following
types
$$\matrix
& & V^{AB} \;\; {\ssize (spinor)} & & \\
& \nearrow & \updownarrow & \searrow &                                   \\
V^i & \leftrightarrow & V^I & \leftrightarrow & V^\mu  \\
& \searrow & \updownarrow & \nearrow & \\
& & V^I \;\; {\ssize(complex)} & &
\endmatrix$$ and the stroke covariant derivative of a vector or tensor quantity
transforms in the appropriate way.

Let $\bold{\Psi}$ be a 4-spinor with components $\Psi^A$.  Let $D_\mu$ represent the
stroke covariant derivative operator.  We conjecture that the full Lagrangian is given
by
$$ \aligned
\Cal{L} = & i\alpha \overline{\bold{\Psi}} \bold{\gamma^\mu} D_\mu \bold{\Psi} + C^\mu
C_\mu h
\\ = & \alpha i \Psi_A^\dagger (\gamma^0)^A_B (\gamma^\mu)^B_C \Psi^C_{\;\; |\mu}+ C^\mu C_\mu h
\endaligned \tag{40}
$$
where $\alpha$ is an arbitrary real constant and the stroke derivative is given by
(38).  This Lagrangian is invariant under all conservative coordinate transformations
and all differentiable frame transformations. If $\Bbb A^a_C$ is constant and if
$C_\mu=0$, the Lagrangian reduces to that of a free particle of spin 1/2.  As the
transformations allowed in this new geometry includes local Lorentz transformations,
local complex Lorentz transformations, local spin transformations and conservative
transformations on Greek indices, we suggest that the geometry has sufficient richness
to describe the unification of gravitational, electroweak and strong forces.

\par \phantom{D}

{\bf Acknowledgments.}  Many thanks to Professor Dave Pandres who began this research
effort and who helped me through many discussions over the years.  Also, the author
thanks Greg Cook, Paul Anderson and Wake Forest University for helpful discussions and
their hospitality at the 8th East Coast Gravity Meeting.

 \heading References \endheading

\flushpar Bade, W. L. and Jehle, H.: An introduction to spinors. Rev. Mod. Phys. 25,
714-728 (1953)
\newline Barut, A. O.: Electrodynamics and Classical Theory of Fields
and Particles, 1st ed. Dover,  New York (1980)
\newline Clarke, C. J. S. and de Felice, F.:  Relativity on Curved Manifolds. Cambridge
University Press, Cambridge (1992)
\newline Dirac, P. A. M.: The Principles of quantum Mechanics. Cambridge University Press,
Cambridge (1930)
\newline  Green, E. L. and Pandres, D., Jr.: Unified field theory from enlarged
transformation group. The consistent Hamiltonian.  Int. J. Theor. Phys. 42, 1849-1873
(2003)
\newline Kibble, T. W. B.: Lorentz invariance and the gravitational field. J. Math.
Phys. 2, 212-221 (1961)
\newline Hehl, F. W., von der Heyde, H., Kerlick, G. D., Nester, J. M.:
General relativity with spin and torsion:  Foundations and prospects. Rev. Mod. Phys.
48, 393-416 (1976)
\newline Pandres, D., Jr.: Quantum unified field theory from enlarged coordinate
transformation group. Phys. Rev. D 24, 1499-1508 (1981)
\newline Pandres, D., Jr.:  Quantum unified field theory from enlarged coordinate
transformation group. II. Phys. Rev. D 30, 317-324 (1984)
\newline Ryder, L. H.: Quantum Field Theory, 2nd ed. Cambridge University
Press, Cambridge (1996)
\newline Weinberg, S.: Gravitation and Cosmology. Wiley, New York (1972)

\end